\documentclass[cameraready]{Interspeech}

\title{Stuttering Classification and Segmentation with Attention-Based Multiple Instance Learning}

\author[affiliation={1},orcid=0009-0003-5741-7616]{Petar}{Sušac}
\author[affiliation={2},orcid=0000-0002-3502-9511]{Sebastian P.}{Bayerl}
\author[affiliation={1},orcid=0000-0003-0502-6923]{Hrvoje}{Džapo}

\address{
    $^1$ University of Zagreb Faculty of Electrical Engineering and Computing, Croatia \\
    $^2$ Rosenheim Technical University of Applied Sciences, Germany
}

\email{\{petar.susac,hrvoje.dzapo\}@fer.unizg.hr, sebastian.bayerl@th-rosenheim.de}

\usepackage{comment}
\usepackage{subcaption}
\usepackage{amsmath}
\usepackage{microtype}

\begin{document}

\maketitle

\begin{abstract}
    Stuttering detection and classification using deep learning methods has the
    potential to improve the process of stuttering severity assessment. Most
    stuttering classification datasets provide clip-level labels, making them
    unsuitable for fine-grained frame-level classification needed to determine
    the duration of individual stuttering dysfluencies. To overcome this
    challenge, we present a multiple instance neural network architecture based
    on fine-tuned wav2vec 2.0, WavLM and Whisper encoders. We apply instance-
    and embedding-based multiple instance learning approaches to train
    models on a clip-level dataset for both clip-level and frame-level
    stuttering classification tasks. Our results show a 23\% improvement in
    frame-level F1 score and between 2\% and 9\% in clip-level F1 score,
    demonstrating the ability of our models to utilize clip-level data for
    frame-level segmentation.
\end{abstract}

\section{Introduction}

Stuttering is a speech fluency disorder characterized by involuntary
dysfluencies such as blocks, repetitions and prolongations that disrupt the
natural flow of speech. The research of stuttering detection and classification
using machine learning methods has gained popularity in recent years due to its
potential to automate the process of stuttering severity assessment and
monitoring, as well as to improve the experience of using speech recognition
technology for people who stutter \cite{alnashwan_computational_2023a,
sheikh_machine_2022a, barrett_systematic_2022}. The goal of stuttering
detection/classification systems is to detect stuttering dysfluencies in speech
recordings and classify them into dysfluency types (blocks, prolongations,
repetitions, etc.) Stuttering classification can be performed at the clip level,
where the goal is to determine the presence or absence of a dysfluency type in
an audio clip, or at the frame level, where the goal is to determine the exact
timestamps of the dysfluencies.

Several stuttered speech datasets in various languages have been created to
support these efforts. Some of these datasets are annotated with dysfluency
timestamps \cite{valente_clinical_2025, batra_boli_2025a}, but most are
annotated at the clip level, where the exact timestamps of the dysfluencies
within the clips are unknown \cite{lea_sep28k_2021, bayerl_ksof_2022a,
gong_as70_2024}. The clip-level labeling method is considered to be more
practical as it can be applied by non-expert annotators and allows them to label
a large amount of data more quickly \cite{bayerl_ksof_2022a}. However, when
assessing the intensity of stuttering in a clinical setting, the exact duration
of dysfluencies is important. Some of the standardized stuttering severity
assessment instruments like the SSI-4 \cite{riley_stuttering_1972} and the
speech efficiency score (SES) \cite{amir_speech_2018} require the duration of
dysfluencies to be known. Labeling the stuttering events in the form of
fixed-length clips, while practical, neglects this important information.

Due to the limited availability of frame-level annotations, stuttering
classification methods most often focus on clip-level classification, which is
treated as a multi-label classification task due to the fact that more than one
stuttering dysfluency may be present within a clip \cite{bayerl_stutter_2023a}.
Recent methods often rely on foundation speech encoders like wav2vec 2.0
\cite{bayerl_stutter_2023a, bayerl_detecting_2022, miyahara_stuttering_2025,
sen_comparative_2024}, Whisper \cite{changawala_whister_2024,
batra_exploring_2025} or WavLM \cite{shih_selfsupervised_2024}.  To obtain a
clip-level classification decision, the temporal dimension of the encoder output
is reduced using mean pooling \cite{bayerl_detecting_2022,
miyahara_stuttering_2025} or dot-product attention \cite{bayerl_stutter_2023a}
before feeding the hidden representations into the classification head. Other
classification methods, utilizing hand-engineered spectral features (e.g.,
spectrograms, MFCC, pitch) with classifier models such as Long Short-Term Memory
(LSTM) \cite{jouaiti_dysfluency_2022a, narasinga_enhancing_2025} or random
forest (RF) \cite{batra_boli_2025a, jouaiti_dysfluency_2022b}, are also
represented.

Frame-level stuttering classification methods, although less common, have also
been developed. YOLO-Stutter \cite{zhou_yolostutter_2024a} is a multimodal
speech/text model trained on an artificially created stuttering dataset, which
performs multi-class frame-level stuttering classification by formulating the
task as a YOLO-like object detection problem. StutterCut
\cite{ghosh_stuttercut_2025} is a graph clustering method aided by a
Whisper-based clip-level classifier, which performs frame-level stuttering
segmentation by clustering the frames of a clip into fluent and dysfluent
clusters. Harvill et al. \cite{harvill_framelevel_2022} and Shih et al.
\cite{shih_selfsupervised_2024} perform binary frame-level stuttering
classification by pre-training a classifier on synthetic dysfluencies and then
fine-tuning on a clip-level dataset using the multiple-instance learning (MIL)
paradigm.

We improve upon current state-of-the-art (SOTA) methods in both clip-level and
frame-level stuttering classification by generalizing the existing MIL-based
methods to the multi-label classification setting, and by applying the
attention-based embedding MIL approach to this task for the first time to
further improve frame-level performance. Unlike previous methods, our method
does not require pre-training on frame-based datasets and is capable of
zero-shot frame-level classification while only being trained on clip-level
datasets. Our contributions are:
\begin{enumerate}
    \item A multiple-instance neural network (MINN) model architecture achieving
    SOTA clip-level multi-label stuttering classification results on the
    SEP-28k-E dataset, 
    \item Achieving SOTA frame-level stuttering classification performance on
    the CASA annotations of the FluencyBank dataset.
\end{enumerate}

\section{Method}
\subsection{Multiple instance learning}

Frame-level stuttering classification can be learned from clip-level data by
formulating the clip-level stuttering classification task as a weakly-supervised
MIL task. Under this formulation, each audio clip is divided into a number of
frames. In the context of MIL, a frame represents an instance, and a clip is a
collection of frames called a bag of instances. Under the standard MIL
assumption \cite{carbonneau_multiple_2018}, the label of the bag is known, but
the labels of individual instances are unknown. If any instance within a bag has
a positive label, then the label of the bag will also be positive. If all of the
instance labels within a bag are negative, then the bag label will also be
negative. In the context of multi-label stuttering classification, a clip is
labeled with a dysfluency label if at least one of its frames contains the
dysfluency.

Deep neural networks can be applied to MIL tasks using one of two methods: the
instance-based and the embedding-based approach \cite{wang_revisiting_2018}.
When the instance-based approach is used, the network outputs a label
probability for each instance. The instance-level probabilities are aggregated
into a bag-level probability using a MIL pooling function. The max, mean or the
log-sum-exp functions are commonly used for this purpose
\cite{wang_revisiting_2018}. In this case, the instance scores provide
interpretability to the bag-level classification decision - instances with a
high positive class probability contribute towards a positive classification
decision. On the other hand, the embedding-based approach maps a collection of
instances to a bag embedding, which is then used to infer a bag-level label.
This approach is considered to result in better bag classification performance
\cite{wang_revisiting_2018}, however it lacks the interpretability of the
instance-based approach. The attention-based MIL pooling operator
\cite{ilse_attentionbased_2018} was introduced to bridge the two approaches. It
allows a model to learn an adaptive pooling function which corresponds to a
version of the attention mechanism. The attention weights of the operator have
been shown to identify key instances that contribute to a positive bag-level
classification in embedding-based MIL models \cite{ilse_attentionbased_2018}.

Existing MIL methods of stuttering classification utilize the instance-based
approach and apply max pooling to aggregate instance-level predictions into a
bag-level prediction \cite{shih_selfsupervised_2024, harvill_framelevel_2022}.
The attention pooling embedding-based approach has not yet been applied to
stuttering classification, but it has been applied to analyze the physiological
signals of children who stutter to investigate which signal features could be
used to identify stuttering events \cite{sharma_psychophysiological_2022a}.

\begin{figure*}[t]
    \centering
    \begin{subfigure}{0.5\linewidth}
        \centering
        \includegraphics[trim={0 1cm 0 0.5cm},clip,height=6cm]{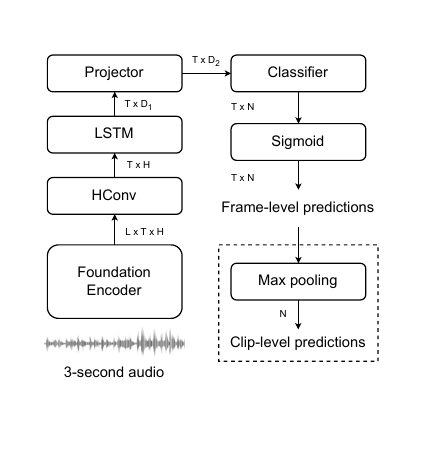}
        \caption{Instance-based model}
        \label{fig:inst_model}
    \end{subfigure}%
    \begin{subfigure}{0.5\linewidth}
        \centering
        \includegraphics[trim={0 1cm 0 0.25cm},clip,height=6cm]{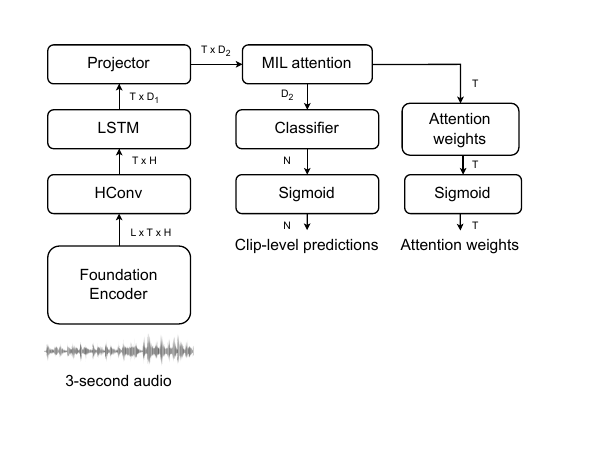}
        \caption{Embedding-based model}
        \label{fig:attn_model}
    \end{subfigure}
    \caption{Architecture of the proposed MINN models. The tensor dimensions are: $L$ = number of encoder layers, $T$ = number of frames (temporal dimension), $H$ = encoder embedding size, $D_1, D_2$ = LSTM/projector embedding size, $N$ = number of labels}
    \label{fig:models}
\end{figure*}

\begin{figure}[h!]
    \centering
    \begin{subfigure}{\linewidth}
        \centering
        \includegraphics[width=\linewidth]{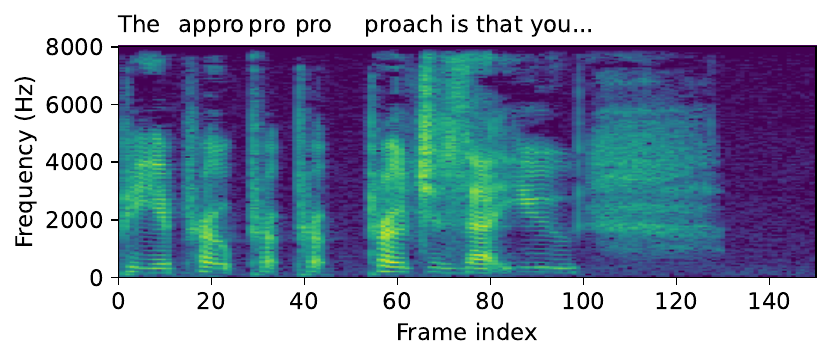}
        \caption{Spectrogram and transcription of a stuttered speech sample}
        \vspace{1mm}
        \label{fig:spectrogram}
    \end{subfigure}
    \begin{subfigure}{\linewidth}
        \centering
        \includegraphics[width=\linewidth]{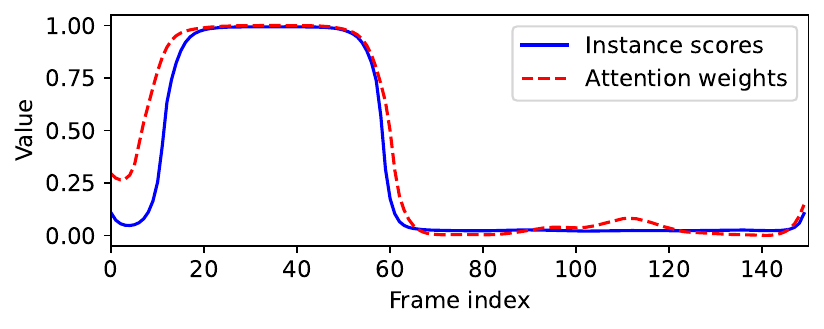}
        \caption{Instance-based model's instance scores (solid blue) and embedding-based model's attention weights (dashed red). The attention weights are rescaled as $a_i' = (a_i - \min(\mathbf{a})) / (\max(\mathbf{a}) - \min(\mathbf{a}))$}
        \label{fig:frame_weights}
    \end{subfigure}
    \caption{Spectrogram and single-label frame-level model outputs for a clip
    from the SEP-28k-E test set. Both models used the Whisper encoder}
    \label{fig:frame_weights_example}
\end{figure}

\subsection{Model architecture}

We use two MINN models in our experiments: an instance-based model utilizing
max-pooling and an embedding-based model utilizing MIL attention pooling. Their
architecture is shown in Figure \ref{fig:models}. The first stage of both models
is a pre-trained foundation encoder. Drawing inspiration from previous work, we
tested wav2vec 2.0 \cite{baevski_wav2vec_2020a}, Whisper
\cite{radford_robust_2023} and WavLM \cite{chen_wavlm_2022a} encoders and based
our models on their pretrained checkpoints: \texttt{wav2vec2-large},
\texttt{whisper-medium}, and \texttt{wavlm-large}, respectively. The checkpoints
were chosen due to their comparable number of parameters (wav2vec 2.0: 315M,
WavLM: 315M, Whisper: 307M) and embedding size (1024). Following the approach by
Shih et al. \cite{shih_selfsupervised_2024}, we use the HConv interface
\cite{shih_interface_2024} to pool the outputs of multiple encoder layers. The
convolutional feature extractors of the encoders produce $T$ frame-level
representations of the input clip. The duration of a frame is 20 ms for all
three foundation encoders ($T=150$ frames per 3-second input clip). The essence
of our MIL approach is to structure the rest of the network in a way that
preserves the temporal dimension and allows the network to learn frame-level
labels. The outputs of the encoder are passed through a bidirectional LSTM
network consisting of 4 layers of 512 units. The LSTM acts as a smoothing layer
for the frame-level outputs, ensuring temporal consistency between frames which
leads to improved frame-level performance. The LSTM outputs are further passed through a
projector consisting of 2 layers of 256 and 128 fully connected neurons with a
leaky ReLU non-linearity.

For the instance-based model, the output of the projector is passed through an
instance-based multi-label classification head with a sigmoid activation. The
instance-based classifier returns a multi-label vector of classification
probabilies for each instance. To obtain the clip-level classification result,
the maximum value of the instance probabilities for each label is selected. When
using the instance-based model for frame-level segmentation, the max-pooling
layer is removed and a threshold $\theta$ is applied to the outputs of the
instance-based classifier.

For the embedding-based model, the output of the projector is passed through a
MIL attention layer. We follow the MIL attention operator structure introduced
by Ilse et al. \cite{ilse_attentionbased_2018}, which is implemented by two
fully-connected layers of 128 and $T$ neurons, along with a $\tanh$ nonlinearity
and a softmax function. The produced attention weights are of size $T$. The bag
representation created by applying the attention-pooling operator is passed to a
multi-label classification head. When performing frame-level segmentation
inference, if the clip is labeled as negative, then all of the frames of the
clip are considered to be negative. If the clip is labeled as positive, the
unnormalized attention weights (before applying the softmax function) are used
for frame-level classification by applying the sigmoid activation function and a
threshold $\theta$. The rationale behind using the attention weight values
before applying the softmax function is to prevent the influence of the duration
of a dysfluent region on the weight values. Since the softmax function makes the
weights sum to 1, the individual normalized weight values would be lower if the
total sum of the unnormalized weights were greater due to a prolonged
dysfluency. An example of computed instance probabilities and attention weights
for a stuttered speech sample is shown in Figure
\ref{fig:frame_weights_example}.

\subsection{Loss function}

The models are trained using the binary cross-entropy (BCE) loss function,
taking the mean loss value across labels for multi-label models. To handle class
imbalance, we apply a weighting factor to the positive samples for each label:
\begin{equation}
    w^+_l = \frac{N^-_l}{N^+_l},
\end{equation}
where $l$ is the label, and $N^-_l$ and $N^+_l$ are the
number of negative and positive samples for label $l$, respectively. In
addition, we apply a per-sample weighting factor based on annotator agreement on
the \emph{No stuttered words} label. The labels are described in more detail in section \ref{sec:data}. Each sample is labeled by 3 annotators.
Samples that are not unanimously agreed on are weighted with a factor of 0.25: 
\begin{equation}
    \begin{split}
        w_i = \begin{cases}
            1 & \mathit{NS}(i) \in \{0, 3\} \\
            0.25 & \mathit{NS}(i) \in \{1, 2\} \\
        \end{cases}
    \end{split}
\end{equation}
where $\mathit{NS}(i)$ is the number of annotator votes for the \emph{No
stuttered words} label of the $i$-th batch sample. The annotator agreement
weights are further normalized per batch so that the mean weight across the
batch is equal to 1:
\begin{equation}
    \begin{split}
        w_\mathit{i, norm} = \frac{B}{\sum_{j=1}^{B} w_j} w_i\\
    \end{split}
\end{equation}
where $B$ is the batch size.

\section{Experiments}

\subsection{Data} \label{sec:data}

We train our models on the standardized SEP-28k-E split of the clip-level
SEP-28k dataset \cite{bayerl_influence_2022}. The dataset contains 28,000
3-second clips labeled with the following dysfluency labels: \emph{Block},
\emph{Prolongation}, \emph{Sound repetition}, \emph{Word repetition},
\emph{Interjection}, and \emph{No stuttered words}. Each sample was labeled by 3
annotators. We consider a label positive if at least 2 annotators voted for the
label. We invert the \emph{No stuttered words} label to comply with the MIL
assumption (a positive label means a dysfluency is present). To evaluate
cross-dataset performance and compare with the current SOTA MINN model
\cite{shih_selfsupervised_2024}, we also report clip-level performance on the
clip-level labels of the FluencyBank dataset \cite{lea_sep28k_2021}, comprised
of 4,144 clips annotated using the same labels as the SEP-28k. Since the
baseline model is a single-label binary classifier, for this purpose we train a
single-label classifier using only the inverted \emph{No stuttered words} label.

Frame-level performance is evaluated on the timestamped CASA
annotations~\cite{valente_clinical_2025} of the FluencyBank dataset. These
annotations consist of core (vocal) and secondary (visual) behaviors, marked by
3 annotators. The core labels are: \emph{Syllable repetition}, \emph{Incomplete
syllable repetition}, \emph{Multisyllable unit repetition}, \emph{Sound
prolongation}, and \emph{Block}. We evaluate on the "gold standard" consensus
test set containing 8 recordings of variable length with a total of 732
dysfluencies \cite{valente_clinical_2025}. Due to the differences in labels
between SEP-28k and CASA, only the single-label performance is evaluated by
aggregating the primary dysfluency labels of the CASA dataset. The recordings
are divided into 3-second clips with no overlap to be used as the input of our
models. The clips are segmented into 20 ms frames and frame labels are created
based on the dysfluency timestamps.

\subsection{Setup}

For our training, we use the Adam optimizer with an initial learning rate of $5
\times 10^{-5}$ and a batch size of 16. First, the foundation encoder is frozen
and the model is trained until the development set loss stops decreasing for 3
epochs. Then, the foundation encoder is unfrozen and fine-tuned with an initial
learning rate of $1 \times 10^{-5}$ until the development set loss stops
decreasing again. The model with the lowest development set loss across all
training epochs is evaluated on the test set. For the clip-level task, we apply
a threshold of $\theta = 0.5$ to the classifier output. For the frame-level
task, we compare the labels to the outputs of the instance-based max-pooling
classifier or the unnormalized attention weights of the embedding-based
classifier, with a threshold $\theta = 0.5$ in both cases.

\subsection{Results}

\begin{table}[ht]
  \caption{Clip-level multi-label F1 scores on the SEP-28k-E test set (Bl: Block,
  Pro: Prolongation, Snd: Sound repetition, Wd: Word repetition, Int: Interjection,
  NS: No stuttered words)}
  \label{tab:clip_results}
  \centering
  \footnotesize
  \tabcolsep=0.15cm
  \begin{tabular}{lcccccc}
    \toprule
    \textbf{Model} & \textbf{Bl} & \textbf{Pro} & \textbf{Snd} & \textbf{Wd} & \textbf{Int} & \textbf{NS} \\
    \midrule
    Miyahara et al. \cite{miyahara_stuttering_2025} & 0.30 & \textbf{0.53} & 0.46 & 0.67 & 0.78 & \textbf{0.82} \\
    Haas et al. \cite{haas_multilingual_2026} & 0.33 & 0.51 & \textbf{0.53} & 0.71 & 0.77 & - \\
    \midrule
    wav2vec 2.0 + max. pool & 0.30 & 0.25 & 0.46 & 0.34 & 0.40 & 0.70 \\
    wav2vec 2.0 + attn. pool & 0.31 & 0.39 & 0.44 & 0.41 & 0.71 & 0.69 \\
    WavLM + max. pool & 0.34 & 0.31 & 0.51 & 0.72 & \textbf{0.83} & 0.78 \\
    WavLM + attn. pool & \textbf{0.35} & 0.47 & 0.42 & 0.74 & 0.82 & 0.78 \\
    Whisper + max. pool & \textbf{0.35} & 0.50 & \textbf{0.53} & \textbf{0.80} & 0.74 & 0.78 \\
    Whisper + attn. pool & \textbf{0.35} & 0.49 & \textbf{0.53} & 0.78 & 0.82 & 0.78 \\
    \bottomrule
  \end{tabular}
\end{table}

\begin{table}[ht]
    \caption{Clip-level single-label results on the FluencyBank dataset}
    \label{tab:fluencybank_results}
    \centering
    \footnotesize
    \begin{tabular}{lccc}
        \toprule
        \textbf{Model} & \textbf{F1} & \textbf{Precision} & \textbf{Recall} \\
        \midrule
        Shih et al. \cite{shih_selfsupervised_2024} & 0.85 & - & - \\
        \midrule
        wav2vec 2.0 + max. pool & 0.74 & 0.75 & 0.72 \\
        wav2vec 2.0 + attn. pool & 0.77 & 0.86 & 0.70 \\
        WavLM + max. pool & 0.88 & 0.88 & 0.89 \\
        WavLM + attn. pool & 0.89 & 0.89 & 0.88 \\
        Whisper + max. pool & 0.89 & 0.86 & 0.92 \\
        Whisper + attn. pool & \textbf{0.90} & 0.91 & 0.89\\
        \bottomrule
    \end{tabular}
\end{table}

\begin{table}[ht]
    \caption{Frame-level single-label results on the "gold standard" test set of the CASA dataset}
    \label{tab:boli_results}
    \centering
    \footnotesize
    \begin{tabular}{lccc}
        \toprule
        \textbf{Model} & \textbf{F1} & \textbf{Precision} & \textbf{Recall}\\
        \midrule
        YOLO-Stutter \cite{zhou_yolostutter_2024a} & 0.47 & 0.47 & 0.49 \\
        StutterCut \cite{ghosh_stuttercut_2025} & 0.45 & 0.39 & 0.58 \\
        \midrule
        WavLM + max. pool & 0.46 & 0.53 & 0.41 \\
        WavLM + attn. pool & 0.56 & 0.53 & 0.59 \\
        Whisper + max. pool & 0.66 & 0.76 & 0.59 \\
        Whisper + attn. pool & \textbf{0.70} & 0.71 & 0.69\\
        \bottomrule
    \end{tabular}
\end{table}

The clip-level multi-label classification results are reported in Table
\ref{tab:clip_results}. We compare our results with two baselines that achieved
SOTA multi-label classification results on the SEP-28k-E split without any
samples being excluded from the dataset: a VGG-19 classifier on self-attention
weights of wav2vec 2.0 features by Miyahara et al.
\cite{miyahara_stuttering_2025} and a wav2vec 2.0-based model trained on
multilingual data by Haas et al. \cite{haas_multilingual_2026}. The clip-level
results of our single-label models on the FluencyBank dataset, compared to the
current SOTA instance-based MINN model utilizing a WavLM encoder with
max-pooling \cite{shih_selfsupervised_2024}, are shown in Table
\ref{tab:fluencybank_results}. The baseline scores are quoted directly from
their papers.

We compare the frame-level performance of our models on the CASA dataset to
YOLO-Stutter \cite{zhou_yolostutter_2024a} and StutterCut
\cite{ghosh_stuttercut_2025}, the SOTA stuttering segmentation methods. For
YOLO-Stutter, we obtain audio transcriptions with WhisperX
\cite{bain_whisperx_2023} and use the inference code provided by the authors.
For StutterCut, we use our own implementation. We evaluate our WavLM- and
Whisper-based models due to their superior performance to the wav2vec 2.0-based
models on the clip-level task. We calculate the frame-level F1 score, precision
and recall for each recording, and report the average across all recordings in
Table \ref{tab:boli_results}.

\section{Discussion}

Our WavLM- and Whisper-based models achieve SOTA results in the clip-level
detection of blocks, sound repetitions, word repetitions and interjections. Our
results fall slightly short of the baselines for prolongations and the general
dysfluent label. The improvement in results might depend most on the
architecture of the foundational encoders, considering our wav2vec 2.0-based
models underperform the other variants. Our models also achieved an improvement
on the cross-dataset single-label classification task when compared to the
baseline instance-based MIL model. Here, the baseline model used the same WavLM
encoder as some of our models, but its weights were frozen during training, so
we attribute our improvement to the fine-tuning stage, which reduced the need
for pretraining on artificial stuttering. Our attention-based embedding approach
resulted in a slight improvement over instance-based models on this dataset,
demonstrating its usefulness on the clip-level task.

Our frame-level results indicate a substantial frame-level F1 improvement
compared to the baselines. However, it should be noted that the baseline methods
assume every input clip contains a dysfluency, which is not the case in our
evaluation on whole speech recordings, resulting in some false positives. Unlike
the baselines, our models do not require any prior knowledge of the presence of
dysfluencies, making them more suitable for end-to-end processing. Both Whisper
and WavLM embedding-based models achieved a higher F1 score than instance-based
ones and showed improved recall, which is especially important in clinical
diagnostic tasks. This result further demonstrates the effectiveness of the
embedding-based approach on this task over existing instance-based methods.
Through manual examination of the frame-level predictions, we have noticed that
our models sometimes struggle with long-lasting blocks. This is likely caused by
the fact that the models can only make predictions based on a 3-second context
window, which might lack the necessary acoustic evidence in the case of blocks
that span multiple window lengths. Longer context windows might be necessary to
further improve the performance on this type of dysfluency.

\section{Conclusion}

Our work investigates the application of the weakly-supervised multiple instance
learning paradigm to the task of stuttering classification, being the first to
explore MIL for multi-label stuttering classification, as well as the
application of attention-pooling embedding-based MINNs to this task. We discover
that instance-based and embedding-based MINN classifiers, taking advantage of
speech foundation encoders, can be effectively used for clip-level multi-label
stuttering classification. Additionally, the advantage of MINN models over the
current SOTA clip-level classifiers is a level of interpretability that could
prove useful in clinical applications. Our models achieve multi-label
classification results comparable to SOTA models on the SEP-28k-E dataset. The
same models demonstrate SOTA performance on frame-level stuttered speech
segmentation. The embedding-based MIL approach has proved to be a viable option
for utilizing datasets with clip-level labels to train models for this task,
demonstrating superior performance over the current best segmentation methods.
To further improve the frame-level performance, pre-training or fine-tuning on a
frame-level dataset might be useful, as demonstrated by some previous studies
\cite{shih_selfsupervised_2024, harvill_framelevel_2022}. Multi-label
frame-level segmentation performance should also be evaluated in future studies.
Our results show that the MIL paradigm could be a promising approach towards
clinically applicable stuttering assessment algorithms by allowing existing
clip-level datasets to be used for training frame-level classifiers, which are
necessary for the automated assessment of stuttering severity with standardized
instruments.

\section{Acknowledgements}
This research was supported by the European Union-\mbox{NextGenerationEU} project NPOO
581-16956 VISTAHealth. Computing resources were provided by the University of
Zagreb Computing Centre (SRCE) through the Advanced Computing service.

\section{Generative AI Use Disclosure}
Generative AI tools (Claude Sonnet 4.6) were used to assist with grammar and
spelling, with all changes reviewed and approved by the authors.

\bibliographystyle{IEEEtran}
\bibliography{bibliography}

\begin{thebibliography}{10}
\providecommand{\url}[1]{#1}
\csname url@samestyle\endcsname
\providecommand{\newblock}{\relax}
\providecommand{\bibinfo}[2]{#2}
\providecommand{\BIBentrySTDinterwordspacing}{\spaceskip=0pt\relax}
\providecommand{\BIBentryALTinterwordstretchfactor}{4}
\providecommand{\BIBentryALTinterwordspacing}{\spaceskip=\fontdimen2\font plus
\BIBentryALTinterwordstretchfactor\fontdimen3\font minus
  \fontdimen4\font\relax}
\providecommand{\BIBforeignlanguage}[2]{{%
\expandafter\ifx\csname l@#1\endcsname\relax
\typeout{** WARNING: IEEEtran.bst: No hyphenation pattern has been}%
\typeout{** loaded for the language `#1'. Using the pattern for}%
\typeout{** the default language instead.}%
\else
\language=\csname l@#1\endcsname
\fi
#2}}
\providecommand{\BIBdecl}{\relax}
\BIBdecl

\bibitem{alnashwan_computational_2023a}
R.~Alnashwan, N.~Alhakbani, A.~{Al-Nafjan}, A.~Almudhi, and W.~{Al-Nuwaiser},
  ``Computational {{Intelligence-Based Stuttering Detection}}: {{A Systematic
  Review}},'' \emph{Diagnostics}, vol.~13, no.~23, p. 3537, Nov. 2023.

\bibitem{sheikh_machine_2022a}
S.~A. Sheikh, M.~Sahidullah, F.~Hirsch, and S.~Ouni, ``Machine learning for
  stuttering identification: {{Review}}, challenges and future directions,''
  \emph{Neurocomputing}, vol. 514, pp. 385--402, Dec. 2022.

\bibitem{barrett_systematic_2022}
L.~Barrett, J.~Hu, and P.~Howell, ``Systematic {{Review}} of {{Machine Learning
  Approaches}} for {{Detecting Developmental Stuttering}},'' \emph{IEEE/ACM
  Transactions on Audio, Speech, and Language Processing}, vol.~30, pp.
  1160--1172, 2022.

\bibitem{valente_clinical_2025}
A.~Valente, R.~Marew, H.~Toyin, H.~{Al-Ali}, A.~Bohnen, I.~Becerra, E.~Soares,
  G.~Leal, and H.~Aldarmaki, ``Clinical {{Annotations}} for {{Automatic
  Stuttering Severity Assessment}},'' in \emph{Interspeech 2025}.\hskip 1em
  plus 0.5em minus 0.4em\relax ISCA, Aug. 2025, pp. 4318--4322.

\bibitem{batra_boli_2025a}
A.~Batra, M.~Narang, N.~K. Sharma, and P.~K. Das, ``Boli: {{A}} dataset for
  understanding stuttering experience and analyzing stuttered speech,'' in
  \emph{{{IEEE International Conference}} on {{Acoustics}}, {{Speech}} and
  {{Signal Processing}} ({{ICASSP}})}, Apr. 2025, pp. 1--4.

\bibitem{lea_sep28k_2021}
C.~Lea, V.~Mitra, A.~Joshi, S.~Kajarekar, and J.~P. Bigham, ``{{SEP-28k}}: {{A
  Dataset}} for {{Stuttering Event Detection}} from {{Podcasts}} with {{People
  Who Stutter}},'' in \emph{{{IEEE International Conference}} on {{Acoustics}},
  {{Speech}} and {{Signal Processing}} ({{ICASSP}})}, Jun. 2021, pp.
  6798--6802.

\bibitem{bayerl_ksof_2022a}
S.~Bayerl, A.~{Wolff von Gudenberg}, F.~H{\"o}nig, E.~Noeth, and K.~Riedhammer,
  ``{{KSoF}}: {{The Kassel State}} of {{Fluency Dataset}} -- {{A Therapy
  Centered Dataset}} of {{Stuttering}},'' in \emph{Proceedings of the
  {{Thirteenth Language Resources}} and {{Evaluation Conference}}}.\hskip 1em
  plus 0.5em minus 0.4em\relax European Language Resources Association, Jun.
  2022, pp. 1780--1787.

\bibitem{gong_as70_2024}
R.~Gong, H.~Xue, L.~Wang, X.~Xu, Q.~Li, L.~Xie, H.~Bu, S.~Wu, J.~Zhou, Y.~Qin,
  B.~Zhang, J.~Du, J.~Bin, and M.~Li, ``{{AS-70}}: {{A Mandarin}} stuttered
  speech dataset for automatic speech recognition and stuttering event
  detection,'' in \emph{Interspeech 2024}.\hskip 1em plus 0.5em minus
  0.4em\relax ISCA, Sep. 2024, pp. 5098--5102.

\bibitem{riley_stuttering_1972}
G.~D. Riley, ``A {{Stuttering Severity Instrument}} for {{Children}} and
  {{Adults}},'' \emph{Journal of Speech and Hearing Disorders}, vol.~37, no.~3,
  pp. 314--322, Aug. 1972.

\bibitem{amir_speech_2018}
O.~Amir, Y.~Shapira, L.~Mick, and J.~S. Yaruss, ``The {{Speech Efficiency
  Score}} ({{SES}}): {{A}} time-domain measure of speech fluency,''
  \emph{Journal of Fluency Disorders}, vol.~58, pp. 61--69, Dec. 2018.

\bibitem{bayerl_stutter_2023a}
S.~P. Bayerl, D.~Wagner, I.~Baumann, F.~H{\"o}nig, T.~Bocklet, E.~N{\"o}th, and
  K.~Riedhammer, ``A {{Stutter Seldom Comes Alone}} -- {{Cross-Corpus
  Stuttering Detection}} as a {{Multi-label Problem}},'' in \emph{Interspeech
  2023}, 2023, pp. 1538--1542.

\bibitem{bayerl_detecting_2022}
S.~P. Bayerl, D.~Wagner, E.~Noeth, and K.~Riedhammer, ``Detecting
  {{Dysfluencies}} in {{Stuttering Therapy Using}} wav2vec 2.0,'' in
  \emph{Interspeech 2022}.\hskip 1em plus 0.5em minus 0.4em\relax ISCA, Sep.
  2022.

\bibitem{miyahara_stuttering_2025}
G.~Miyahara, T.~Kato, and A.~Tamura, ``Stuttering {{Detection Based}} on
  {{Self-Attention Weights}} of {{Temporal Acoustic Vector Sequence}},'' in
  \emph{Interspeech 2025}.\hskip 1em plus 0.5em minus 0.4em\relax ISCA, Aug.
  2025, pp. 5298--5302.

\bibitem{sen_comparative_2024}
M.~Sen, A.~Batra, and P.~K. Das, ``Comparative {{Analysis}} of {{Classifiers}}
  using {{Wav2Vec2}}.0 {{Layer Embeddings}} for {{Imbalanced Stuttering
  Datasets}},'' in \emph{International {{Conference}} on {{Electronics}},
  {{Communication}} and {{Signal Processing}} ({{ICECSP}})}, Aug. 2024, pp.
  1--6.

\bibitem{changawala_whister_2024}
V.~Changawala and F.~Rudzicz, ``Whister: {{Using Whisper}}'s representations
  for {{Stuttering}} detection,'' in \emph{Interspeech 2024}.\hskip 1em plus
  0.5em minus 0.4em\relax ISCA, Sep. 2024, pp. 897--901.

\bibitem{batra_exploring_2025}
A.~Batra, B.~Kar, and P.~K. Das, ``Exploring {{Whisper Embeddings}} for
  {{Stutter Detection}}: {{A Layer-Wise Study}},'' \emph{33rd European Signal
  Processing Conference (EUSIPCO 2025)}, 2025.

\bibitem{shih_selfsupervised_2024}
Y.-J. Shih, Z.~Gkalitsiou, A.~G. Dimakis, and D.~Harwath, ``Self-{{Supervised
  Speech Models For Word-Level Stuttered Speech Detection}},'' in \emph{{{IEEE
  Spoken Language Technology Workshop}}, {{SLT}}}.\hskip 1em plus 0.5em minus
  0.4em\relax IEEE, 2024, pp. 937--944.

\bibitem{jouaiti_dysfluency_2022a}
M.~Jouaiti and K.~Dautenhahn, ``Dysfluency {{Classification}} in {{Stuttered
  Speech Using Deep Learning}} for {{Real-Time Applications}},'' in
  \emph{{{IEEE International Conference}} on {{Acoustics}}, {{Speech}} and
  {{Signal Processing}} ({{ICASSP}})}, May 2022, pp. 6482--6486.

\bibitem{narasinga_enhancing_2025}
V.~Narasinga, P.~Kommagouni, S.~Vanga, K.~S.~S. Motepalli, S.~Akarsh~C,
  P.~Barche, and A.~Vuppala, ``Enhancing {{Stutter Detection}} using
  {{Long-Term Average Spectrum Values}},'' in \emph{{{IEEE International
  Conference}} on {{Acoustics}}, {{Speech}} and {{Signal Processing}}
  ({{ICASSP}})}, Apr. 2025, pp. 1--5.

\bibitem{jouaiti_dysfluency_2022b}
M.~Jouaiti and K.~Dautenhahn, ``Dysfluency {{Classification}} in {{Speech
  Using}} a {{Biological Sound Perception Model}},'' in \emph{9th
  {{International Conference}} on {{Soft Computing}} \& {{Machine
  Intelligence}} ({{ISCMI}})}, Nov. 2022, pp. 173--177.

\bibitem{zhou_yolostutter_2024a}
X.~Zhou, A.~Kashyap, S.~Li, A.~Sharma, B.~Morin, D.~Baquirin, J.~Vonk,
  Z.~Ezzes, Z.~Miller, M.~Tempini, J.~Lian, and G.~Anumanchipalli,
  ``{{YOLO-Stutter}}: {{End-to-end Region-Wise Speech Dysfluency Detection}},''
  in \emph{Interspeech 2024}.\hskip 1em plus 0.5em minus 0.4em\relax ISCA, Sep.
  2024, pp. 937--941.

\bibitem{ghosh_stuttercut_2025}
S.~Ghosh, M.~Jouaiti, J.-O. Perschewski, and S.~Stober, ``{{StutterCut}}:
  {{Uncertainty-Guided Normalised Cut}} for {{Dysfluency Segmentation}},'' in
  \emph{Interspeech 2025}, 2025, pp. 808--812.

\bibitem{harvill_framelevel_2022}
J.~Harvill, M.~{Hasegawa-Johnson}, and C.~D. Yoo, ``Frame-{{Level Stutter
  Detection}},'' in \emph{Interspeech 2022}.\hskip 1em plus 0.5em minus
  0.4em\relax ISCA, Sep. 2022, pp. 2843--2847.

\bibitem{carbonneau_multiple_2018}
M.-A. Carbonneau, V.~Cheplygina, E.~Granger, and G.~Gagnon, ``Multiple instance
  learning: {{A}} survey of problem characteristics and applications,''
  \emph{Pattern Recognition}, vol.~77, pp. 329--353, May 2018.

\bibitem{wang_revisiting_2018}
X.~Wang, Y.~Yan, P.~Tang, X.~Bai, and W.~Liu, ``Revisiting multiple instance
  neural networks,'' \emph{Pattern Recognition}, vol.~74, pp. 15--24, Feb.
  2018.

\bibitem{ilse_attentionbased_2018}
M.~Ilse, J.~Tomczak, and M.~Welling, ``Attention-based {{Deep Multiple Instance
  Learning}},'' in \emph{Proceedings of the 35th {{International Conference}}
  on {{Machine Learning}}}.\hskip 1em plus 0.5em minus 0.4em\relax PMLR, Jul.
  2018, pp. 2127--2136.

\bibitem{sharma_psychophysiological_2022a}
H.~Sharma, Y.~Xiao, V.~Tumanova, and A.~Salekin, ``Psychophysiological
  {{Arousal}} in {{Young Children Who Stutter}}: {{An Interpretable AI
  Approach}},'' \emph{Proceedings of the ACM on Interactive, Mobile, Wearable
  and Ubiquitous Technologies}, vol.~6, no.~3, pp. 137:1--137:32, Sep. 2022.

\bibitem{baevski_wav2vec_2020a}
A.~Baevski, Y.~Zhou, A.~Mohamed, and M.~Auli, ``Wav2vec 2.0: {{A Framework}}
  for {{Self-Supervised Learning}} of {{Speech Representations}},'' in
  \emph{Advances in {{Neural Information Processing Systems}}}, vol.~33.\hskip
  1em plus 0.5em minus 0.4em\relax Curran Associates, Inc., 2020, pp.
  12\,449--12\,460.

\bibitem{radford_robust_2023}
A.~Radford, J.~W. Kim, T.~Xu, G.~Brockman, C.~Mcleavey, and I.~Sutskever,
  ``Robust {{Speech Recognition}} via {{Large-Scale Weak Supervision}},'' in
  \emph{Proceedings of the 40th {{International Conference}} on {{Machine
  Learning}}}.\hskip 1em plus 0.5em minus 0.4em\relax PMLR, Jul. 2023, pp.
  28\,492--28\,518.

\bibitem{chen_wavlm_2022a}
S.~Chen, C.~Wang, Z.~Chen, Y.~Wu, S.~Liu, Z.~Chen, J.~Li, N.~Kanda,
  T.~Yoshioka, X.~Xiao, J.~Wu, L.~Zhou, S.~Ren, Y.~Qian, Y.~Qian, J.~Wu,
  M.~Zeng, X.~Yu, and F.~Wei, ``{{WavLM}}: {{Large-Scale Self-Supervised
  Pre-Training}} for {{Full Stack Speech Processing}},'' \emph{IEEE Journal of
  Selected Topics in Signal Processing}, vol.~16, no.~6, pp. 1505--1518, Oct.
  2022.

\bibitem{shih_interface_2024}
Y.-J. Shih and D.~Harwath, ``Interface {{Design}} for {{Self-Supervised Speech
  Models}},'' in \emph{Interspeech 2024}.\hskip 1em plus 0.5em minus
  0.4em\relax ISCA, Sep. 2024, pp. 2504--2508.

\bibitem{bayerl_influence_2022}
S.~P. Bayerl, D.~Wagner, E.~N{\"o}th, T.~Bocklet, and K.~Riedhammer, ``The
  {{Influence}} of~{{Dataset Partitioning}} on~{{Dysfluency Detection
  Systems}},'' in \emph{Text, {{Speech}}, and {{Dialogue}}}.\hskip 1em plus
  0.5em minus 0.4em\relax Springer International Publishing, 2022, pp.
  423--436.

\bibitem{haas_multilingual_2026}
F.~Haas and S.~P. Bayerl, ``Multilingual {{Stutter Event Detection}} for
  {{English}}, {{German}}, and {{Mandarin Speech}},'' in \emph{Text,
  {{Speech}}, and {{Dialogue}}}.\hskip 1em plus 0.5em minus 0.4em\relax
  Springer Nature Switzerland, 2026, vol. 16029, pp. 194--206.

\bibitem{bain_whisperx_2023}
M.~Bain, J.~Huh, T.~Han, and A.~Zisserman, ``{{WhisperX}}: {{Time-Accurate
  Speech Transcription}} of {{Long-Form Audio}},'' in \emph{Interspeech
  2023}.\hskip 1em plus 0.5em minus 0.4em\relax ISCA, Aug. 2023, pp.
  4489--4493.

\end{thebibliography}

\end{document}